\documentclass{svjour3}   
\smartqed  % flush right qed marks, e.g. at end of proof
\usepackage{graphicx}
\usepackage{latexsym}
\usepackage{amssymb}
\usepackage{amsmath}
\usepackage{pxfonts}
\journalname{General Relativity and Gravitation}

\begin{document}
\title{Event horizon silhouette: implications to supermassive black holes M87* and SgrA*}
\author{Vyacheslav I. Dokuchaev \and Natalia O. Nazarova \and Vadim P. Smirnov}
\institute{V. I. Dokuchaev \at Institute for Nuclear Research of the Russian Academy of Sciences, \\
60th October Anniversary Prospect 7a, 117312 Moscow, Russia, \\\email{dokuchaev@inr.ac.ru} 
\and
N. O. Nazarova \at Scuola Internazionale Superiore di Studi Avanzati (SISSA), \\ Via Bonomea 265, 34136 Trieste (TS) Italy, \\\email{natalia.nazarova@sissa.it}  
\and V. P. Smirnov \at Moscow Institute of Physics and Technology, \\ 9 Institutskiy per., Dolgoprudny, Moscow Region, 141700 Russia, \\\email{smirnov.vp@phystech.edu} }
\date{\today}
\maketitle

\begin{abstract}
We demonstrate that a dark silhouette of the black hole illuminated by a thin accretion disk and seen by a distant observer is, in fact, a silhouette of the event horizon hemisphere. The boundary of this silhouette is a contour of the event horizon equatorial circle if a thin accretion disk is placed in the black hole equatorial plane. A luminous matter plunging into black hole from different directions provides the observational opportunity for recovering a total silhouette of the invisible event horizon globe. The event horizon silhouette is projected on the celestial sphere within a position of the black hole shadow. A relative position of brightest point in the accretion disk with respect to the position of event horizon silhouette in the image of black hole in the galaxy M87, observed by the Event Horizon Telescope, corresponds to a rather high value of the black hole spin, $a\simeq0.75$.
\end{abstract}

\section{Introduction}
\label{intro}

Black holes are invisible, or, more definitely, the black hole event horizon is invisible due to the infinite red-shifts of photons emitting outwards from the event horizon. Nevertheless, it is possible for a distant observer to see a dark silhouette of the event horizon illuminated by the non-stationary surrounding matter. 

The supermassive black hole Sagittarius A* at the Galactic Center provides a natural physical laboratory for experimental verification of General Relativity and recovering the black hole silhouette \cite{Gillessen09,Meyer12,Johnson15,Chatzopoulos15,FizLab,Johannsen16c,Eckart17,Zhu19,Zakharov19,TuanDo19}. A weak accretion activity and a quiescent emission from this dormant quasar provides a chance for surrounding plasma to be transparent in the vicinity of event horizon. The global Event Horizon Telescope (EHT) network \cite{Fish16,Lacroix13,Kamruddin,Johannsen16,Johannsen16b,Broderick16,Chael16,Kim16,Roelofs17,Doeleman17} and similar projects such as BlackHoleCam \cite{Goddi17} and GRAVITY \cite{GRAVITY18,GRAVITY19} intend to reveal a real silhouette of the supermassive black hole at the Galactic Center.

A visual form of the black hole silhouette depends on the distribution of surrounding luminous mater. 

In the case of a stationary background light, emitted far enough behind the black hole, this silhouette is a black hole shadow. The image of a black hole shadow is the capture cross-section of photons in the strong gravitational field of the black hole. 

Besides a stationary background light there may be a non-stationary luminous matter in immediate vicinity of the black hole event horizon. For example, the compact stars or clouds of hot gas falling onto the black hole, or innermost part of accretion disk adjoining the event horizon. We demonstrate below that a silhouette of the black hole illuminated by a thin accretion disk and seen by a distant observer above the plane of accretion disk is in fact a silhouette of the northern hemisphere of event horizon.

\section{Gravitational lensing by black hole}

We describe here a gravitational lensing of luminous matter by the Kerr black hole with a gravitational mass $m$ and angular momentum $J=ma$ and seen by a distant observer above the black equatorial plane. It is used the classical Boyer--Lindquist coordinate system \cite{BoyerLindquist} with coordinates $(t,r,\theta,\varphi)$ and with units $G=c=1$. Additionally we put $m=1$. In these units a dimensionless radius of the black hole event horizon is $r_{\rm h}=1+\sqrt{1-a^2}$, where a spin parameter of the black hole $0\leq a\leq1$.

Trajectories of particles (geodesics) with a rest mass $\mu$ in the Kerr space-time are determined by three constants of motion: a total energy $E$, a component of angular momentum parallel to symmetry axis (azimuth angular momentum) $L$ and a relativistic Carter constant $Q$, which is related with the non-equatorial motion of particles \cite{Carter68,Chandra}. At the same time, a radial and latitudinal motions of particles are defined by the radial effective potential
\begin{equation}
R(r)=[E(r^2\!+\!a^2)\!-\!La]^2-(r^2\!-\!2r\!
+\!a^2)[\mu^2r^2\!+\!(L\!-\!aE)^2\!+\!Q]
\label{Rr} 
\end{equation}
and the latitudinal effective potential
\begin{equation}
\Theta(\theta)=Q-\cos\!^2\theta[a^2(\mu^2-E^2)+L^2/\sin\!^{2}\theta].
\label{Vtheta} 
\end{equation}
Meantime, the trajectories of photons (null geodesics with $\mu=0$) in the Kerr space-time are determined only by two dimensionless parameters, $\lambda=L/E$ and $q^2=Q/E^2$. These parameters are related with the impact parameters on the celestial sphere $\alpha$ and $\beta$ seen by a distant observer at a given radius $r_0>>r_{\rm h}$ (i.\,e., practically at infinity), at a given latitude $\theta_0$ and at a given azimuth $\varphi_0$ (see \cite{Bardeen73,CunnBardeen73} for more details):
\begin{equation}
\alpha =-\frac{\lambda}{\sin\theta_0}, \quad
\beta = \pm\sqrt{\Theta(\theta_0)},
\label{Shadow} 
\end{equation}
where $\Theta(\theta)$ is from (\ref{Vtheta}).

We choose three values for a black hole spin parameter: $a=0.9982$ for the fast rotating black hole \cite{Thorne74}, $a=0.65$ for moderately fast rotating black hole \cite{Dokuch14} and $a=0$ for non-rotating Schwarzschild black hole as representative examples. In numerical calculations of photon trajectories we follow the formalism by C. T. Cunnungham and J. M. Bardeen \cite{CunnBardeen73} and choose the particular value of $\theta_0\simeq84^\circ\!\!.\,24$, such that $\cos\theta_0=0.1$. This value of $\theta_0$ is suitable to the supermassive black hole in the Milky Way galaxy (see Figs.~5--11).

\section{Black hole shadow}

\begin{figure}
	\includegraphics[width=0.99\textwidth]{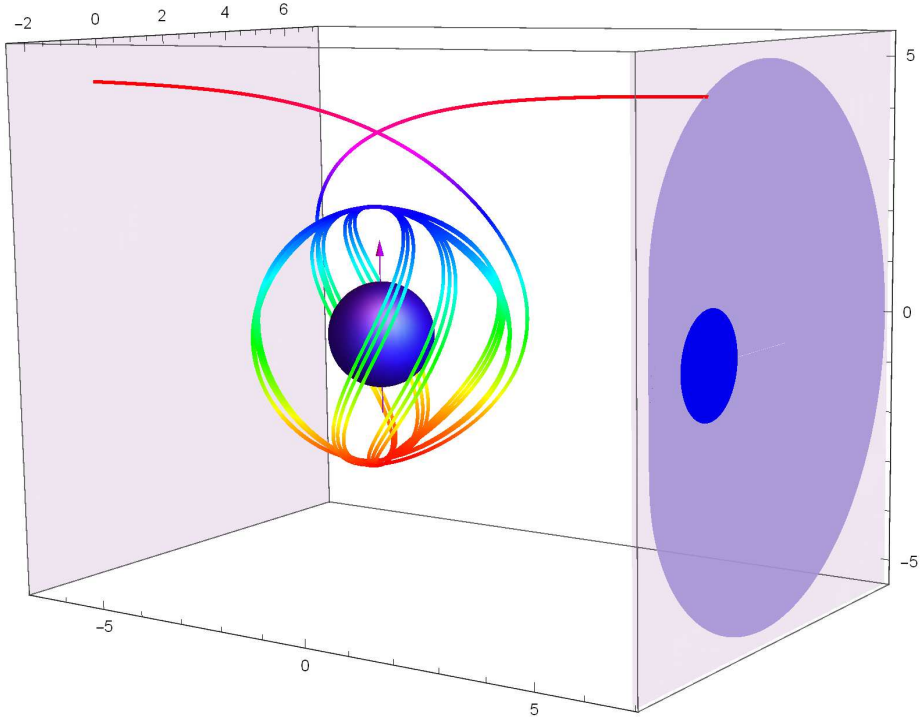}
	\caption{The black hole shadow (dark magenta region) in the case of $a=1$, illuminated by a background (light magenta plane YZ) far behind the black hole with respect to a distant observer in the equatorial plane. It is shown the $3D$ photon trajectory in the Boyer--Lindquist coordinates with orbit parameters $\lambda=0$ and $q=\sqrt{11+8\sqrt{2}}\simeq4.72$, starting from a background behind the black hole and registered by a distant observer on the boundary of black hole shadow, projected on the celestial sphere. A radial turning point is at $r_{\rm min}=1+\sqrt{2}$. A blue sphere is the event horizon globe and, respectively, a blue disk is the event horizon image in the imaginary $3D$ Euclidian space.}
	\label{fig:1}      
\end{figure}
\begin{figure}
	\includegraphics[width=0.47\textwidth]{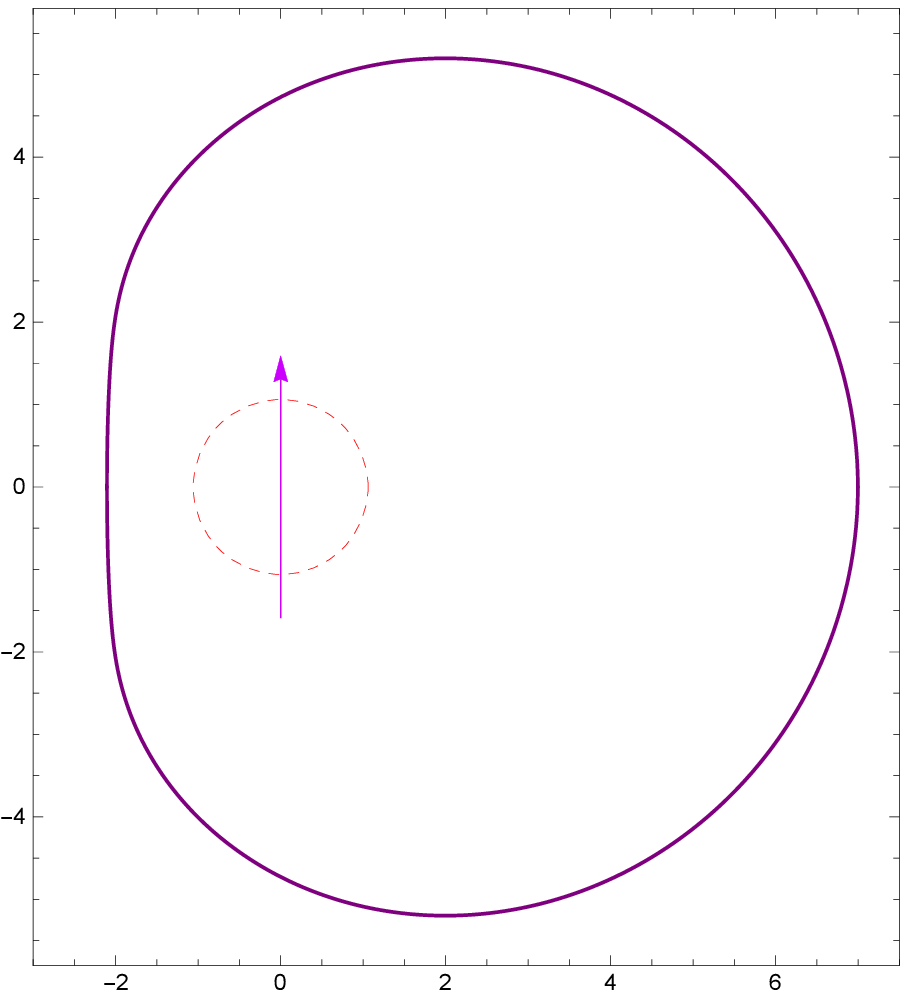} 
	\hskip0.4cm
	\includegraphics[width=0.49\textwidth]{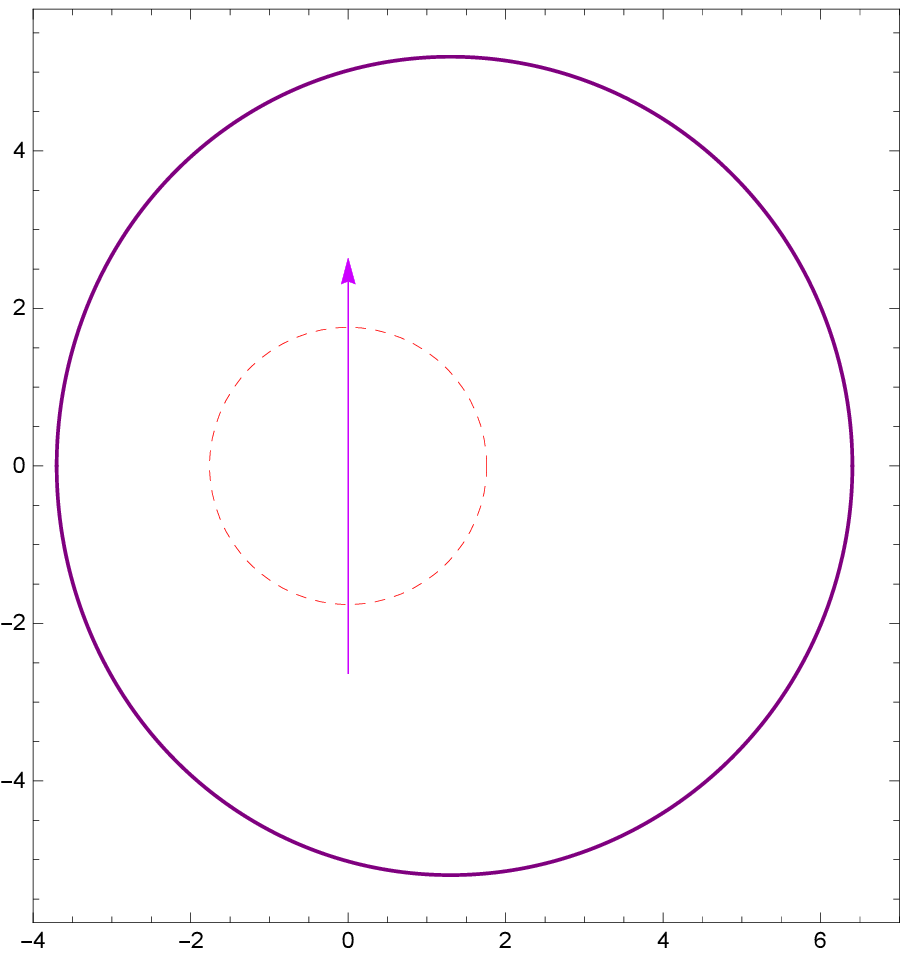}
	\caption{Apparent shape of the Kerr black hole shadows (closed purple curves) for spin parameter $a=0.9982$ (left image) and  $a=0.65$ (right image) seen by a distant observer in the equatorial plane. Dashed magenta circles are the event horizon images in the imaginary $3D$ Euclidian space. Magenta arrows indicate the position of black hole rotation axis.}
	\label{fig:2}  
\end{figure}

A black hole shadow in the Kerr metric, projected on the celestial sphere and seen by a distant observer in the equatorial plane of the black hole, is determined from the simultaneous solution of equations $R(r)=0$ and $[rR(r)]'=0$, where the effective radial potential $R(r)$ is from Eq.~(\ref{Vr}). The corresponding solution for the black hole shadow (for a distant observer in the black hole equatorial plane) in the parametric form $(\lambda,q)=(\lambda(r),q(r))$ is
\begin{eqnarray} \label{shadow1}
\lambda&=&\frac{-r^3+3r^2-a^2(r+1)}{a(r-1)}, \\
q^2&=&\frac{r^3[4a^2-r(r-3)^2]}{a^2(r-1)^2}.
\label{shadow2}
\end{eqnarray}
(see, e.\,g., \cite{Bardeen73,Chandra} for more details). A black hole shadow is the gravitational capture cross-section of photons from the stationary luminous background behind the black holes with respect to the position of a distant observer. In fact, the luminous background should be placed at radial distance from black hole exceeding the radius of photon circular orbit $r_{\rm ph}$ (see definition of $r_{\rm ph}$ in \cite{BPT,Wilkins}) to form the image of the black hole shadow. See in Figs.~1 and 2 the apparent shape of the black hole shadow in the case of a fast rotating black hole with $a=0.9982$ and in the case of black hole with moderate rotation, $a=0.65$. In the spherically symmetric non-rotating case with $a=0$ the radius of black hole shadow is $r_{\rm sh} =3\sqrt{3}\simeq5.196$.

\section{Innermost part of thin accretion disk}

\begin{figure*}
	\includegraphics[width=0.99\textwidth]{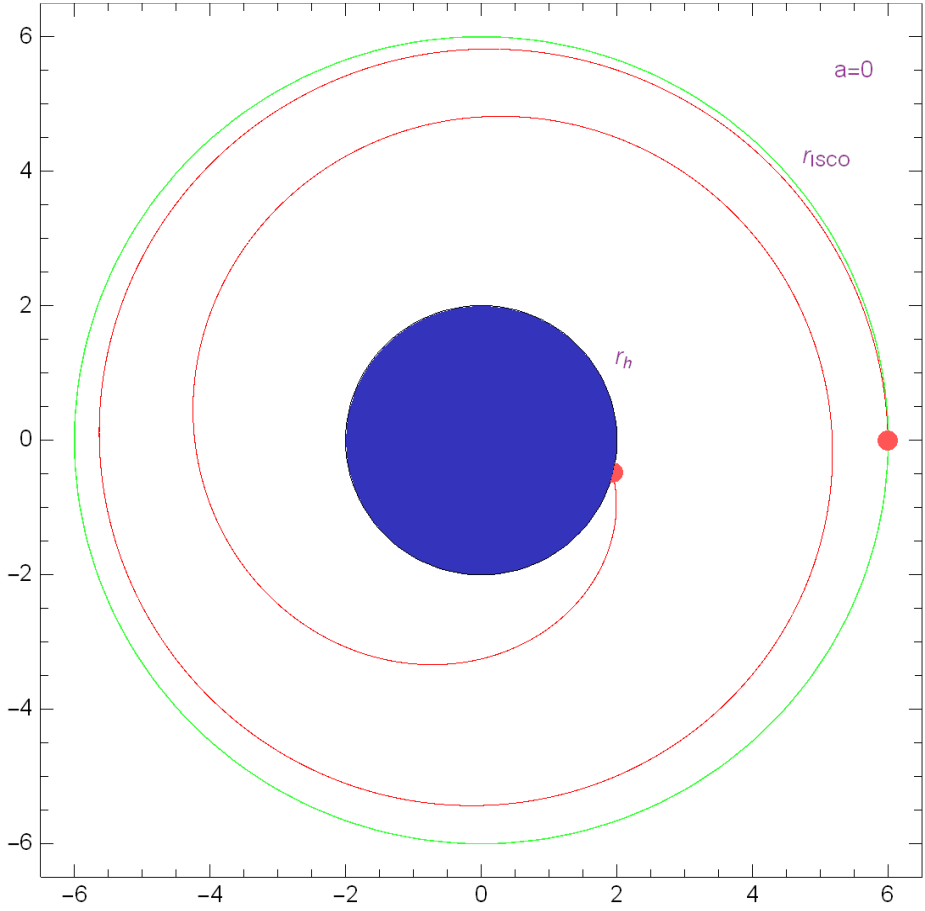}
	\caption{$2D$ infall trajectory of the compact gas cloud (clump) onto the black hole with $a=0$ in the region $r_{\rm h}\leq r\leq r_{\rm ISCO}$. The compact gas cloud is starting at $r=r_{\rm ISCO}$ with orbital parameters $E/\mu=E(r_{\rm ISCO})/\mu=2\sqrt{2}/3$ and $L/\mu=L(r_{\rm ISCO})/\mu-0.001=\sqrt{3}/2-0.001$, where $E$ and $L$ are from (\ref{Ecirc}) and (\ref{Lcirc}).}
	\label{fig:3}      
\end{figure*}
\begin{figure}
	\includegraphics[width=0.99\textwidth]{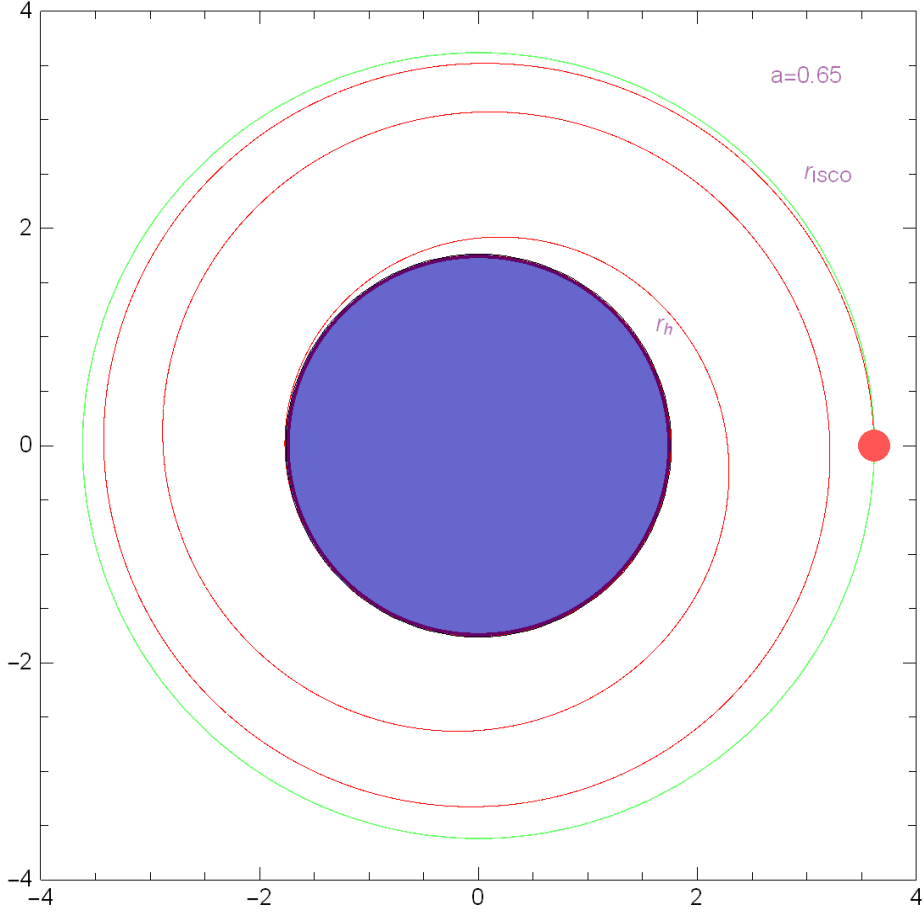}
	\caption{$2D$ infall trajectory of the compact gas cloud (clump) in the equatorial plane of black hole with $a=0.65$ in the region $r_{\rm h}\leq r\leq r_{\rm ISCO}$. The compact gas cloud is starting at $r=r_{\rm ISCO}$ with orbital parameters $E/\mu=E(r_{\rm ISCO})/\mu=0.903123$ and $L/\mu=L(r_{\rm ISCO})/\mu-0.001=2.67469-0.001$, where $E$ and $L$ are from (\ref{Ecirc}) and (\ref{Lcirc}). In contrast with the non-rotating black hole, the falling compact gas cloud is multiply winding around the rotating black hole by approaching the event horizon at $r=r_{\rm h}$.}
	\label{fig:4}      
\end{figure}

Position of the event horizon silhouette on the celestial sphere is recovered by gravitational lensing of the innermost part of accretion disk adjoining the event horizon. Simultaneous solution of equations $R=0$ and  $dR/dr=0$, where the effective radial potential $R$ is from (\ref{Rr}), gives parameters $E$ and $L$ for particles co-rotating with the black hole at a circular orbit with a radius $r$ in the black hole equatorial plane \cite{BPT}:
\begin{eqnarray} 
E/\mu&=&\frac{r^{3/2}-2r^{1/2}+a}{r^{3/4}(r^{3/2}-3r^{1/2}+2a)^{1/2}} \label{Ecirc} \\
L/\mu&=&\frac{r^2-2ar^{1/2}+a^2}{r^{3/4}(r^{3/2}-3r^{1/2}+2a)^{1/2}}.
\label{Lcirc}
\end{eqnarray}
In a geometrically thin accretion disk with negligible self-gravity, there is an inner boundary  for stable circular motion, named the marginally stable radius or the Inner Stable Circular Orbit (ISCO), $r=r_{\rm ISCO}$ \cite{BPT}: 
\begin{equation}\label{ISCO}
r_{\rm ISCO}=3+Z_2-\sqrt{(3-Z_1)(3+Z_1+2Z_2)},
\end{equation}
where
\begin{equation}\label{Z1}
Z_1=1+(1-a^2)^{1/3}[(1+a)^{1/3}+(1-a)^{1/3}],
\end{equation}
\begin{equation}\label{Z2}
Z_2=\sqrt{3a^2+Z_1^2}.
\end{equation}
The non-stationary motion of accreting matter at $r<r_{\rm ISCO}$ weakly depends on the matter viscosity and governs mainly by the black hole gravitational field. We approximate the motion of small gas elements of accreting matter at $r<r_{\rm ISCO}$ by the geodesic motion of separate compact gas clumps with parameters $E$ and $L$ from (\ref{Ecirc}) and (\ref{Lcirc}), corresponding to the radius $r=r_{\rm ISCO}$. Additionally, we suppose that a thin accretion disk is transparent and the energy flux in the rest frame of small gas elements is conserved during their motion in the region $r_{\rm h}\leq r\leq r_{\rm ISCO}$. These model approximations define the lensed brightness of accretion disk image and, at the same time, do not influence the form of the black hole silhouette. See in Figs.~3 and 4 the $2D$ trajectories of falling matter at $r_{\rm h}\leq r\leq r_{\rm ISCO}$. 

\begin{figure}
	\includegraphics[width=0.47\textwidth]{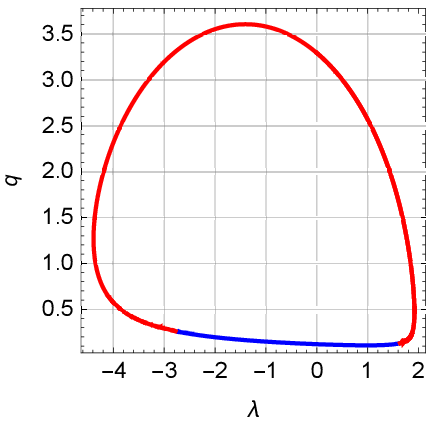} 
	\hskip0.4cm
	\includegraphics[width=0.45\textwidth]{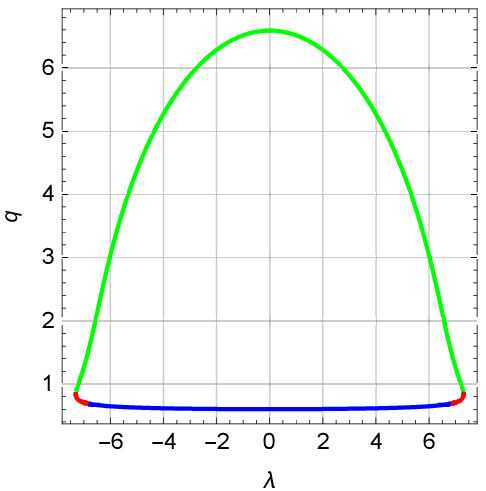}
	\caption{Parameters of photon trajectories $\lambda$ and $q$, reaching a distant observer from compact gas clumps in accretion disk at $r=0.01r_{\rm h}$ (left graph) and, respectively, at $r=r_{\rm ISCO}$ in the case of $a=0$ (right graph). The blue color corresponds to photon trajectories without the turning points, defined from numerical solutions of integral equation (\ref{eq24a}). The red color corresponds to photon trajectories with only one turning point at $\theta=\theta_{\rm min}(\lambda,q)$, defined from numerical solutions of integral equation (\ref{eq24b}). At last, green color corresponds to photon trajectories with two turning points at $\theta=\theta_{\rm min}(\lambda,q)$ and at $r=r_{\rm min}(\lambda,q)$, defined from numerical solutions of integral equation (\ref{eq24c}).}
	\label{fig:5}  
\end{figure}
\begin{figure}
	\includegraphics[width=0.99\textwidth]{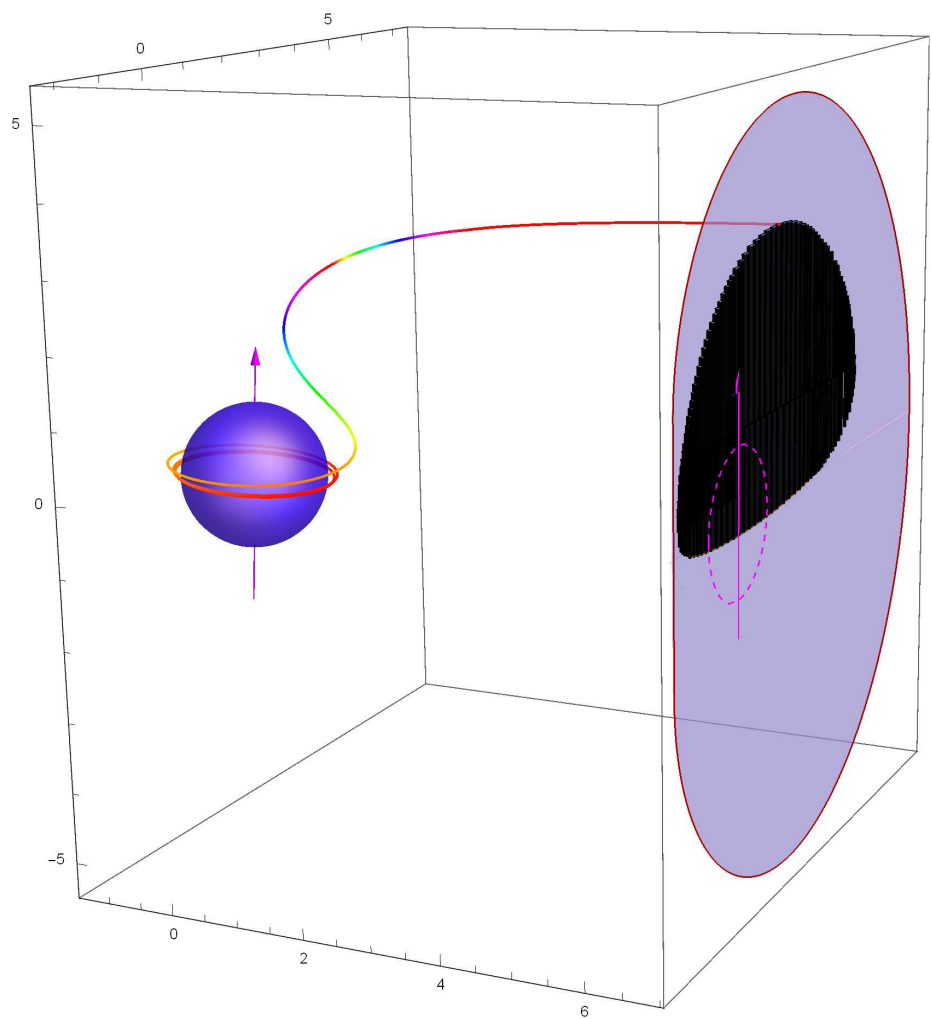}
	\caption{Black hole silhouette (black region) in the case of a thin accretion disk in the equatorial plane of the fast rotating black hole with $a=0.9982$. This silhouette is formed by the lensed photons emitted in the black equatorial plane near the event horizon. These photons are highly red-shifted by reaching a distant observer. It is shown the photon trajectory with $\lambda=0.0632$ and $q=0.1213$, starting in the equatorial plane at $r=1.01r_{\rm h}$ and reaching the distant observer at the boundary of the black hole silhouette. A boundary of this silhouette is a contour of the event horizon equatorial circle. Meantime, the black hole image in this case is a silhouette of the northern hemisphere of the event horizon (see more details in Fig.~11).}
	\label{fig:6}      
\end{figure}
\begin{figure}
	\includegraphics[width=0.95\textwidth]{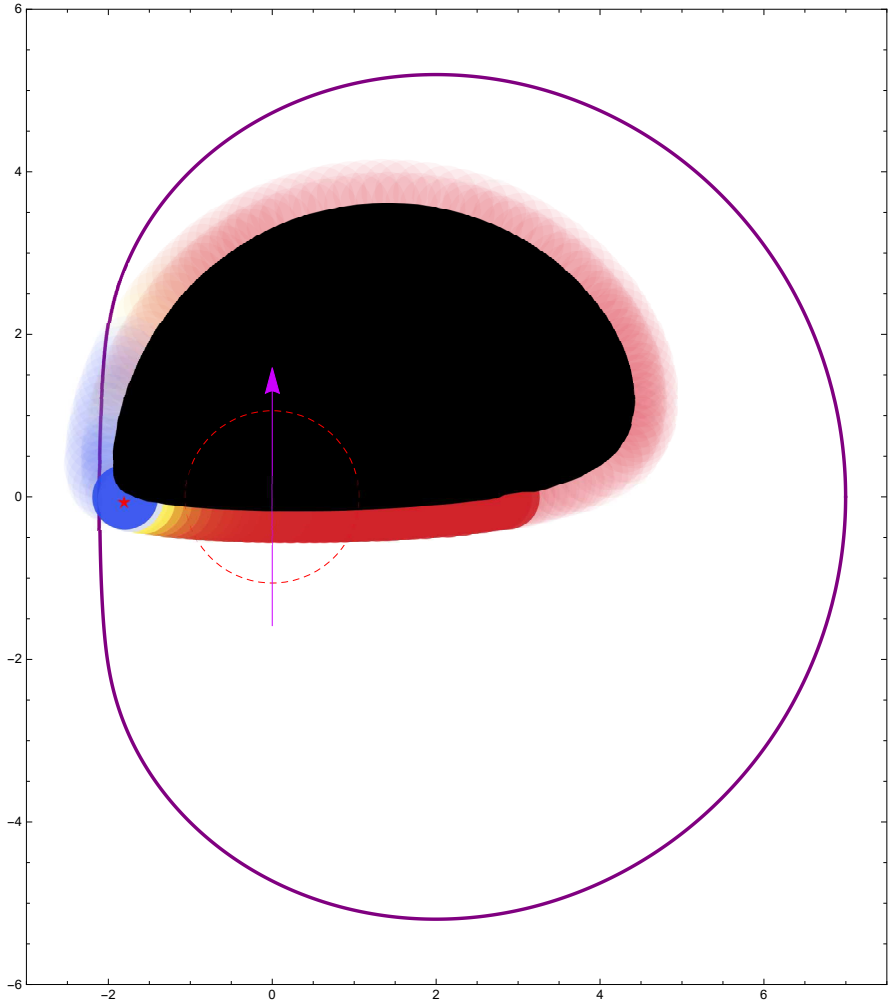}
	\caption{Internal part of the lensed accretion disk at $r_{\rm h}\leq r\leq r_{\rm ISCO}$ adjoining the event horizon in a case of the fast rotating black hole with spin $a=0.9982$. The brightest point (marked by the red star) in  accretion disk is at radius $r_{\rm ISCO}\simeq1.23\simeq1.16r_{\rm h}$ and corresponds to the photon trajectory without turning points, defined from numerical solution of integral equation (\ref{eq24a}), and with the maximum permissible azimuth angular momentum $\lambda=1.71$ ($q=0.144$, $\alpha=-1.72$, $\beta=-0.031$). The black area here and in all similar Figures is the silhouette of the northern hemisphere of black hole event horizon. The contour of this silhouette is the event horizon equatorial circle. A boundary of the black hole shadow is shown by the closed purple curve.}
	\label{fig:7}      
\end{figure}
\begin{figure}
	\includegraphics[width=0.95\textwidth]{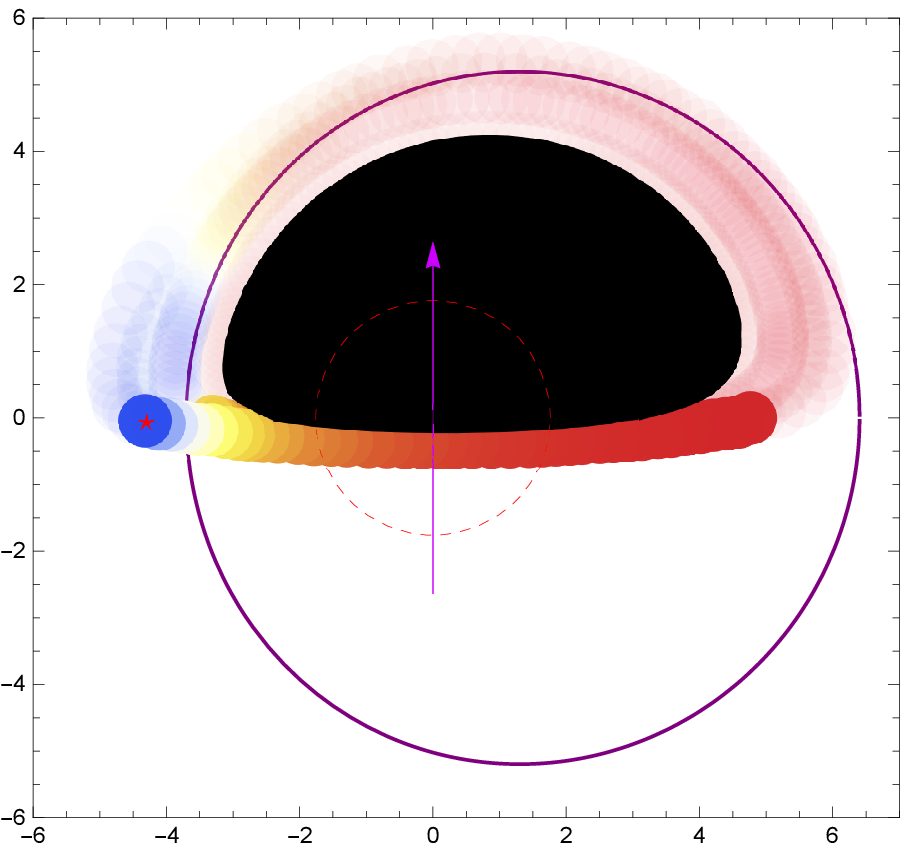}
	\caption{Internal part of the lensed accretion disk at $r_{\rm h}\leq r\leq r_{\rm ISCO}$ adjoining the event horizon in a case of moderately fast rotating black hole with spin $a=0.65$. The brightest point in accretion disk is at radius $r_{\rm ISCO}\simeq3.25\simeq1.85r_{\rm h}$ and corresponds to the photon trajectory with $\lambda=4.29$, $q=0.430$, $\alpha=-4.32$, $\beta=-0.042$.}
	\label{fig:8}      
\end{figure}
\begin{figure}
	\includegraphics[width=0.95\textwidth]{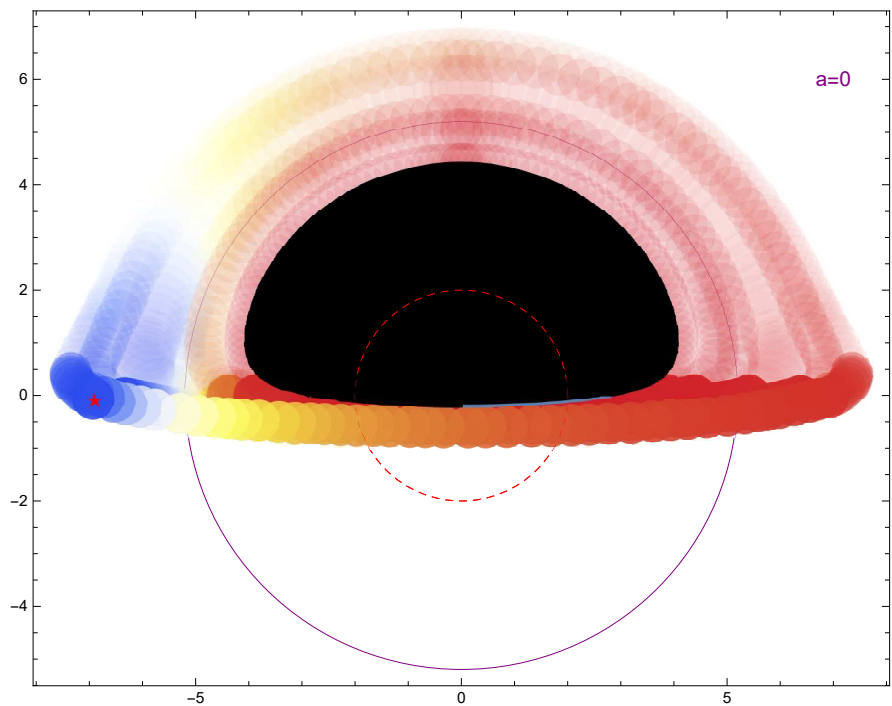}
	\caption{Internal part of the lensed accretion disk at $r_{\rm h}\leq r\leq r_{\rm ISCO}$ adjoining the event horizon in a case of non-rotating black hole with spin $a=0$. The brightest point in accretion disk is at radius $r_{\rm ISCO}=3r_{\rm h}=6$ and corresponds to the photon trajectory with $\lambda=6.89$, $q=0.697$ and $\alpha=-6.92$, $\beta=-0.057$.}
	\label{fig:9}      
\end{figure}
\begin{figure}
	\includegraphics[width=0.29\textwidth]{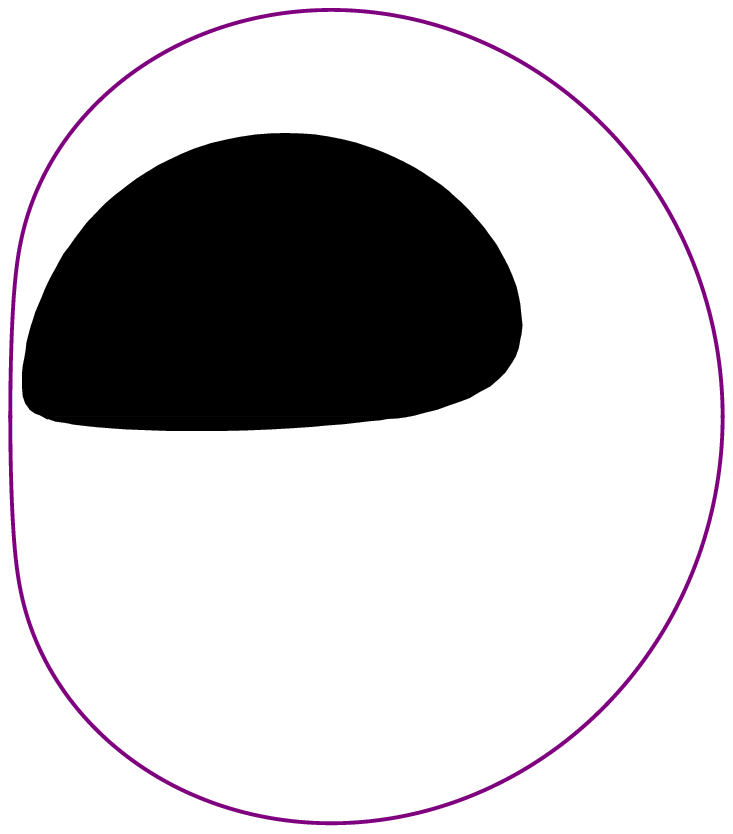}
	\includegraphics[width=0.33\textwidth]{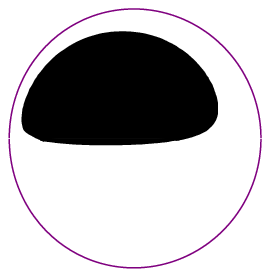}
	\includegraphics[width=0.34\textwidth]{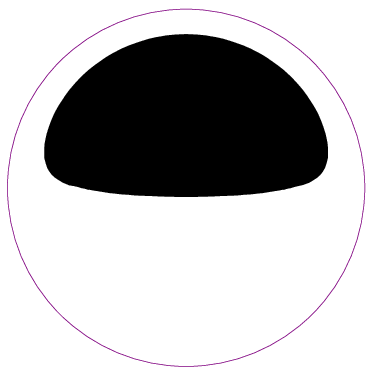}
	\caption{Silhouette of the northern hemisphere of event horizon (black region) placed inside a boundary of the black hole shadow (closed purple curve) for black holes with spin $a=0.9982$, $0.65$ and $0$ (from left to right), viewed by a distance observer with a latitude angle $\theta_0\simeq84^\circ\!\!.\,24$, corresponding to the supermassive black hole in the Milky Way galaxy.}
	\label{fig:10}      
\end{figure}

\section{Energy shift of photons from accretion disk}

We need the expression for energy shift of photons to calculate the energy flux from the lensed image of accretion disk. It is convenient to use the orthonormal Locally Non-Rotating Frames (LNRF) \cite{BPT,Bardeen70}, for which the observers' world lines are $r=const$, $\theta=const$, $\varphi=\omega t+const$,
where a frame dragging angular velocity
\begin{equation}\label{eq2425}
\omega=\frac{2ar}{(r^2+a^2)^2-a^2\Delta\sin\!^{2}\theta},
\end{equation}
and
\begin{equation}\label{eq2425d}
\Delta=r^2-2r+a^2.
\end{equation}
The azimuth velocity at radius $r$ relative the LNRF \cite{BPT,Bardeen70} of a compact gas cloud, falling in the equatorial plane onto a black hole with orbital parameters $E$, $L$ and $Q=0$, is
\begin{equation}\label{eq2425e}
V^{(\varphi)}=\frac{r\sqrt{\Delta}\,L}{[r^3+a^2(r+2)]E-2aL},
\end{equation}
The corresponding radial velocity in the equatorial plane relative the LNRF is
\begin{equation}\label{Vr}
V^{(r)}=-\sqrt{\frac{r^3+a^2(r+2)}{r}}\frac{\sqrt{R(r)}}{[r^3+a^2(r+2)]E-2aL}.
\end{equation}
where $R(r)$ is defined in (\ref{Rr}) with $Q=0$.

We need also the expressions for components of photon 4-momentum in the LNRF:
\begin{equation}\label{p}
p^{(\varphi)}=\sqrt{\frac{r}{r^3+a^2(r+2)}}\,\lambda, \quad 
p^{(t)}=\sqrt{\frac{r^3+a^2(r+2)}{r\Delta}}\,(1-\omega\lambda),
\end{equation}
\begin{equation}\label{pr}
p^{(r)}=-\frac{1}{r}\sqrt{\frac{(r^2+a^2-a\lambda)^2}{\Delta}-[(a-\lambda)^2+q^2]}.
\end{equation}
The photon energy in the LNRF is $E_{\rm LNRF}=p^{(t)}$. At the same time, the photon energy in the orthonormal frame, moving with the velocity $V^{(\varphi)}$ with respect to the LNRF is equal
\begin{equation}\label{EVphi}
E(V^{(\varphi)})=\frac{p^{(t)}-V^{(\varphi)}p^{(\varphi)}}{\sqrt{1-[V^{(\varphi)}]^2}}.
\end{equation}
The energy $E(V^{(\varphi)})$ depends only on $\lambda$. In this frame the compact cloud is still moving with the radial velocity 
\begin{equation}\label{v}
v=\frac{V^{(r)}}{\sqrt{1-[V^{(\varphi)}]^2}}.
\end{equation}
In result, the requested photon energy in the frame, comoving with the compact gas cloud, is equal
\begin{equation}\label{calE}
{\cal{E}}(\lambda,q)=\frac{E(V^{(\varphi)})-vp^{(r)}}{\sqrt{1-v^2}}=\frac{P^{(t)}-V^{(\varphi)}p^{(\varphi)}
	-V^{(r)}p^{(r)}}{\sqrt{1-[V^{(r)}]^2-[V^{(\varphi)}]^2}}.
\end{equation}  
Respectively, the energy shift (the ratio of photon frequency at infinity to the frequency in the rest frame of the compact gas clump) is $g(\lambda,q)=1/{\cal{E}}(\lambda,q)$. We adjust this energy shift to the formalism by C.T.~Cunningham and J.M.~Bardeen \cite{CunnBardeen73} for numerical calculations of the energy flux from the accretion disk measured by a distant observer. The results of these numerical calculations are presented in Figs.~6--12. Local colors of the lensed accretion disk images are related with an effective local black-body temperature, which is proportional to the energy shift $g(\lambda,q)=1/{\cal{E}}(\lambda,q)$. The brightest point in the accretion disk is always placed at radius $r=r_{\rm ISCO}$ and corresponds to the photon trajectory without turning point, defined from numerical solution of integral equation (\ref{eq24a}) and with the maximum permissible azimuth angular momentum $\lambda>0$.

\section{Silhouette of the event horizon}

\begin{figure}
	\includegraphics[width=0.99\textwidth]{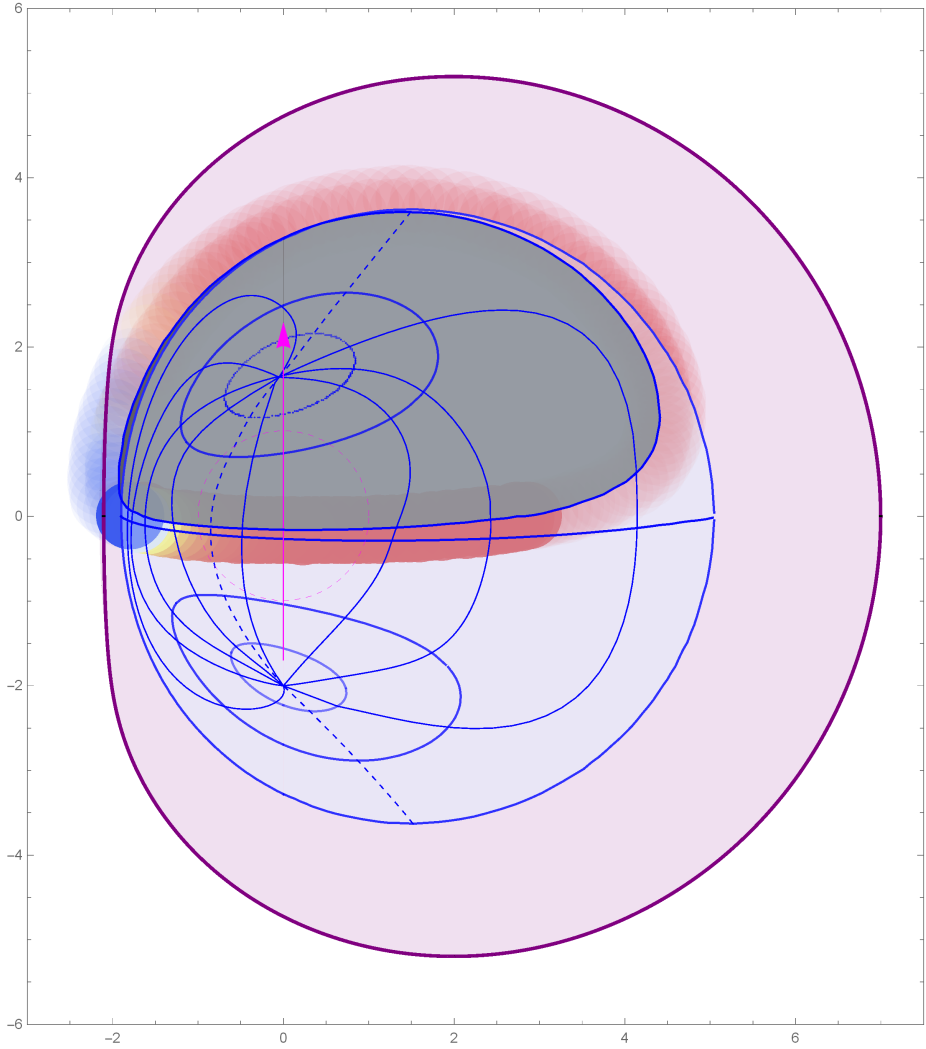}
	\caption{The black hole shadow (magenta region) and silhouette of the northern hemisphere of the event horizon (black region) illuminated by a thin accretion disk in the case of black hole with spin $a=0.9982$. The total globe of the event horizon may be recovered by observations of non-stationary luminous matter plunging into black hole from different directions (see more details in \cite{Dokuch19,Dokuch19b}).}
	\label{fig:11}      
\end{figure}

For numerical calculations of the gravitational lensing by the Kerr black hole we use the integral equations of motion for photons \cite{Carter68,BPT,Chandra}: 
\begin{equation}\label{eq24}
\fint^r\frac{dr}{\sqrt{R(r)}}
=\fint^\theta\frac{d\theta}{\sqrt{\Theta(\theta)}},
\end{equation}
\begin{equation}
\varphi=\!\fint^r\!\frac{a(r^2+a^2-\lambda a)}{(r^2\!-\!2r\!+\!a^2)\sqrt{R(r)}}\,dr
\!+\!\fint^\theta\frac{\lambda\!-\!a\sin^2\theta}{\sin^2\theta\sqrt{\Theta(\theta)}}\,d\theta,
\label{eq25b} %\nonumber
\end{equation}
where the effective potentials $V_r$ and $V_{\rm \theta}$ are from Eqs.~(\ref{Vr}) and (\ref{Vtheta}). The integrals in (\ref{eq24}) and  (\ref{eq25b}) are understood to be path integrals along the trajectory. 

The path integrals in (\ref{eq24}) are the ordinary ones for photon trajectories without the turning points:
\begin{equation}\label{eq24a}
\int^{r_0}_{r_s}\frac{dr}{\sqrt{R(r)}}
=\int^{\theta_0}_{\theta_s}\frac{d\theta}{\sqrt{\Theta(\theta)}}.
\end{equation}
In the case of photon trajectories with one turning point $\theta_{\rm min}(\lambda,q)$ (an extremum of latitudinal potential $\Theta(\theta)$), equation  (\ref{eq24}) is written through the ordinary integrals as
\begin{equation}\label{eq24b}
\int^{r_0}_{r_s}\frac{dr}{\sqrt{R(r)}}
=\int^{\theta_s}_{\theta_{\rm min}}\frac{d\theta}{\sqrt{\Theta(\theta)}}
+\int^{\theta_0}_{\theta_{\rm min}}\frac{d\theta}{\sqrt{\Theta(\theta)}}.
\end{equation}
Respectively, for photon trajectories with two turning points, $\theta_{\rm min}(\lambda,q)$ and $r_{\rm min}(\lambda,q)$  (an extremum of radial potential $R(r)$), equation  (\ref{eq24}) is written through the ordinary integrals as
\begin{equation}\label{eq24c}
\int^{r_r}_{r_{\rm min}}\!\!\!\frac{dr}{\sqrt{R(r)}}
+\!\int^{r_0}_{r_{\rm min}}\!\!\!\frac{dr}{\sqrt{R(r)}}
=\!\!\int^{\theta_s}_{\theta_{\rm min}}\!\!\!\frac{d\theta}{\sqrt{\Theta(\theta)}}
+\!\int^{\theta_0}_{\theta_{\rm min}}\!\!\!\frac{d\theta}{\sqrt{\Theta(\theta)}}.
\end{equation}
Three types of photon trajectories, described by equations (\ref{eq24a}), (\ref{eq24b}) and (\ref{eq24c}), produce the primary image of the accretion disk. Besides the primary image there are also an infinite number of other images (light echoes) of the accretion disk. The energy fluxes from all light echoes are very small in comparison with the energy flux from the primary image. For this reason and for simplicity we describe in the following only the primary image of the accretion disk. See in Fig.~5 the examples of numerical solutions of integral equations  (\ref{eq24a}), (\ref{eq24b}) and (\ref{eq24c}) for the primary images of compact gas clumps at the fixed radius in the accretion disk.

In fact, the event horizon silhouettes, very similar to those shown in Figs.~6--12, were pictured previously in many numerical simulations (see, e.\,g., \cite{Dexter09,Luminet79,Bromley97,Fanton97,Fukue03,Fukue03b,Ru-SenLu16,Luminet19,Shiokawa} for details), however, without the identification of simulated silhouettes with the event horizon hemisphere. 

Note that a total silhouette of the invisible event horizon globe may be recovered by observations of non-stationary luminous matter plunging into a black hole from different directions (see Fig.~11).

\section{Supermassive black hole in the galaxy M87}

\begin{figure}
\includegraphics[width=0.32\textwidth]{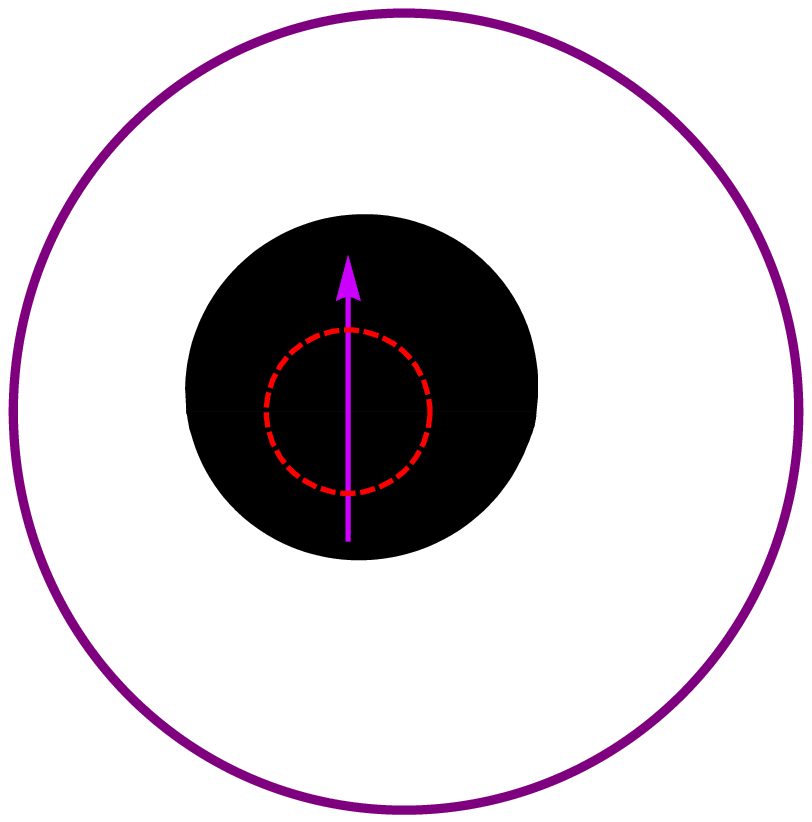}	\includegraphics[width=0.33\textwidth]{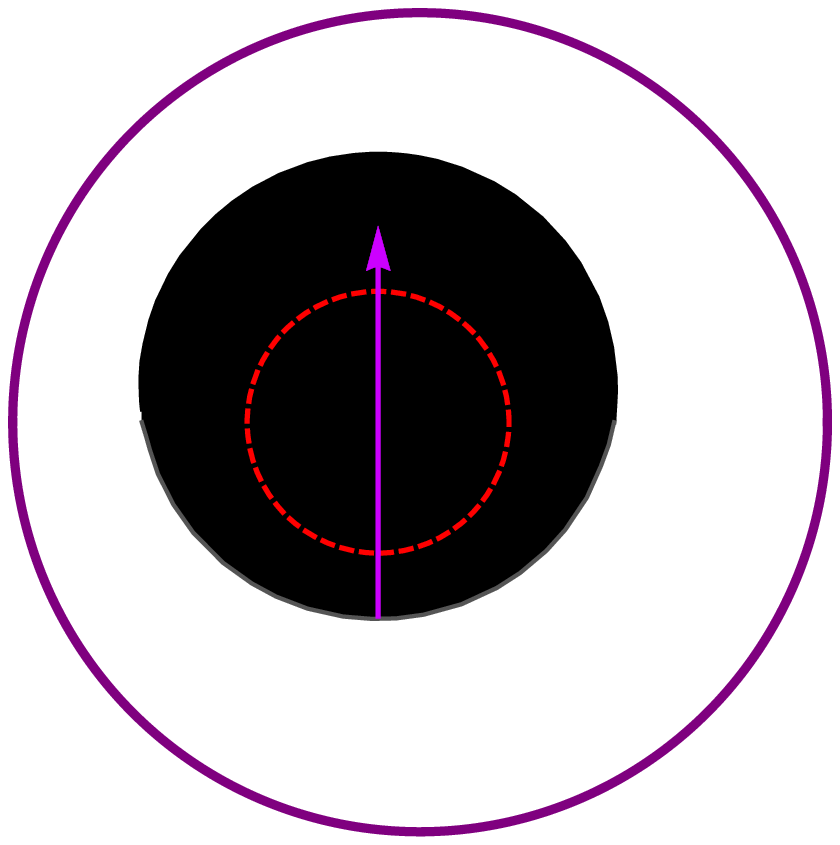}
\includegraphics[width=0.34\textwidth]{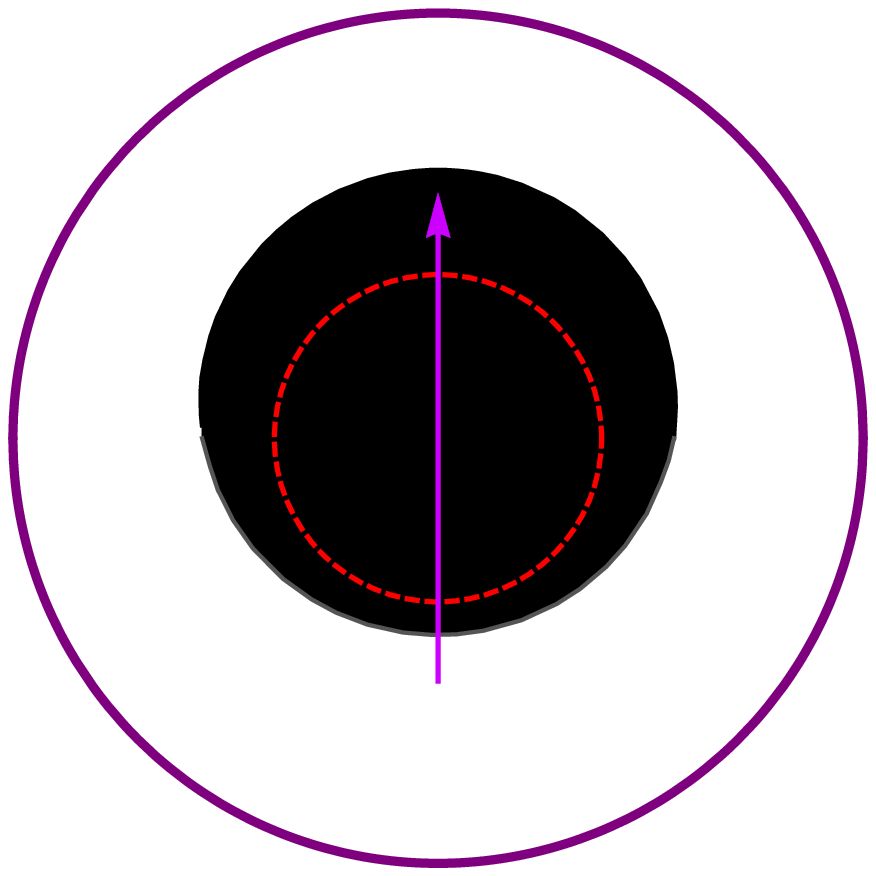}
	\caption{Silhouette of the southern hemisphere of event horizon (black region) placed inside a boundary of the black hole shadow (closed purple curve) for black holes with spin $a=0.9982$, $0.8$ and $0$ (from left to right), viewed by a distance observer with a latitude angle $\theta_0=17^\circ$, corresponding to the optical jet inclination of the  supermassive black hole in the galaxy M87.}
	\label{fig:12}      
\end{figure}

Recently the Event Horizon Telescope consortium presented the first image of the supermassive black hole in the galaxy M87 \cite{EHT1,EHT2,EHT3,EHT4,EHT5,EHT6}. In this image it is viewed the
silhouette of southern hemisphere of event horizon and the bright part of the accretion disk. A black hole shadow is not visible in this image. 

See in Fig.~12 examples of numerically calculated silhouettes of the southern hemisphere of event horizon, projected on the sky plane within the black hole shadow, for black holes with spin $a=0.9982$, $0.8$ and $0$, viewed by a distance observer with a latitude angle $\theta_0=17^\circ$, corresponding to the optical jet inclination of the supermassive black hole in the galaxy M87.

In the limiting case of $\theta_0=0^\circ$ and $a=1$ the black hole shadow is  a circle with radius $r_{\rm sh}=\sqrt{11+8\sqrt{2}}\simeq4.72$, which is less than a corresponding radius $r_{\rm sh}=3\sqrt{3}\simeq5.2$ at $a=0$. 

A relative position of brightest point in the accretion disk with respect to the position of event horizon silhouette in the presented image corresponds to a rather high value of the black hole spin in the galaxy M87, $a\simeq0.75$ (see details in \cite{Dokuch19b}).

\section{Discussions}

A new aspect of our work is in elucidation of difference between the black hole shadow, related with a stationary luminous background behind the black hole, and the silhouette of event horizon, related with the emission of non-stationary matter in the very vicinity of the black hole event horizon (inside the radius of photon circular orbit $r_{\rm ph}$).

A dark silhouette of the black hole event horizon hemisphere is revealed by gravitational lensing of the innermost part of a thin accretion disk adjoining the event horizon, as illustrated in Figs.~6--12. The form of this silhouette is defined only by the black hole gravitational field and does not depend on the used emission model of accretion disk. The brightest point in accretion disk corresponds to the largest (positive) azimuth angular momentum $\lambda$ of photon with the direct orbit, reaching a distant observer without the turning points. 

In the first image of the black hole in the galaxy M87, presented by the Event Horizon Telescope consortium, it is clearly viewed the silhouette of event horizon and the bright part of accretion disk. A black hole shadow is not visible in this image. 

It must be also noted that in numerical simulations for movie ``Interstellar'' it was intentionally neglected by radiation from the non-stationary part of  accretion disk at $r<r_{\rm ISCO}$. For this reason only the black hole shadow is viewed in this movie, not the event horizon silhouette. Additionally, the energy shift of photons, related with the viewed lopsidedness of the accretion disk, in this simulation was also neglected at film producer request \cite{Thorne15,Luminet19b}.

\begin{acknowledgements}
We are grateful to E.O. Babichev, V.A. Berezin, Yu.N. Eroshenko and A.L.~Smir\-nov for stimulating discussions. This work was supported in part by the Russian Foundation for Basic Research grant 18-52-15001a.

\end{acknowledgements}

\end{document}